\documentclass[preprint,12pt]{elsarticle}

\usepackage{amssymb}
\usepackage{amsthm}
\usepackage{booktabs}
\usepackage{subfigure}
\usepackage{amsmath}

\begin{document}

\begin{frontmatter}



% \title{Soft and Hard Constrained Parametric Generative Schemes for Encoding and Synthesizing Airfoils}
\title{Parametric Generative Schemes with Geometric Constraints for Encoding and Synthesizing Airfoils }

\author[inst1]{Hairun Xie}
\author[inst2]{Jing Wang\corref{cor1}}
\ead{wangjinger@sjtu.edu.cn}
\cortext[cor1]{Corresponding author.}

\author[inst1]{Miao Zhang}

\affiliation[inst1]{organization={Shanghai Aircraft Design and Research Institute},%Department and Organization
            % addressline={Rd. Jinke}, 
            city={Shanghai},
            postcode={200436}, 
            % state={State One},
            country={China}}

\affiliation[inst2]{organization={School of Aeronautics and Astronautics, Shanghai Jiao Tong University},%Department and Organization
            % addressline={Address Two}, 
            city={Shanghai},
            postcode={200240}, 
            % state={State Two},
            country={China}}

\begin{abstract}
The modern aerodynamic optimization has a strong demand for parametric methods with high levels of intuitiveness, flexibility, and representative accuracy, which cannot be fully achieved through traditional airfoil parametric techniques.
In this paper, two deep learning-based generative schemes are proposed to effectively capture the complexity of the design space while satisfying specific constraints. 
1. Soft-constrained scheme: a Conditional Variational Autoencoder (CVAE)-based model to train geometric constraints as part of the network directly. 
2. Hard-constrained scheme: a VAE-based model to generate diverse airfoils and an FFD-based technique to project the generated airfoils onto the given constraints.
According to the statistical results, the reconstructed airfoils are both accurate and smooth, without any need for additional filters. 
The soft-constrained scheme generates airfoils that exhibit slight deviations from the expected geometric constraints, yet still converge to the reference airfoil in both geometry space and objective space with some degree of distribution bias. 
In contrast, the hard-constrained scheme produces airfoils with a wider range of geometric diversity while strictly adhering to the geometric constraints. 
The corresponding distribution in the objective space is also more diverse, with isotropic uniformity around the reference point and no significant bias. 
These proposed airfoil parametric methods can break through the boundaries of training data in the objective space, providing higher quality samples for random sampling and improving the efficiency of optimization design.
\end{abstract}

% \includegraphics[width=0.45\textwidth]{"hard_obj.pdf"}

% \begin{highlights}
% \item 
% Two deep learning-based generative schemes are proposed for airfoil synthesis that can effectively capture the complexity of the design space while satisfying specific constraints. The statistical analysis confirms that the reconstructed airfoils are both accurate and smooth, without the need for additional filtering. Moreover, the proposed parametric methods enable the generation of airfoils beyond the boundaries of the training data, which can provide high-quality samples for random sampling and enhance the efficiency of optimization design.
% \end{highlights}

\begin{keyword}
Deep Learning \sep Parameterization Method \sep Parametric Generative Schemes \sep Parametric Airfoil \sep Geometric Constraints  
\end{keyword}

\end{frontmatter}


\section{Introduction}
As the requirements for economy and environmental friendliness of large commercial airliners become increasingly stringent, the geometric parametric technique, allowing for the efficient exploration of the design space and the identification of optimal solutions, should be developed first to meet the increasing demand in engineering design.
The demand for an effective parameterization method is driven by the need for high levels of intuitiveness, flexibility, and representative accuracy. 
Intuitiveness is crucial for designers to build the design logic while flexibility is necessary for accommodating complex design requirements.
Representative accuracy, on the other hand, ensures that a small number of variables can represent a large enough design space, containing optimum aerodynamic shapes for a wide range of design conditions and constraints\cite{kulfan_universal_2008}.

The available parametric methods for aerodynamic design of airfoils typically involve a trade-off between design freedom and intuitiveness\cite{liao_investigation_2011,derksen_bezier-parsec_2010,zhu2011comparison}. 
For instance, the PARSEC approach\cite{sobieczky_parametric_1999} uses explicit mathematical functions to define 2D curves for airfoil, resulting in definitions that are closely tied to geometric features with high intuitiveness. 
However, this approach suffers from relatively limited flexibility and poor representative accuracy, which restricts its application. 
In contrast, the NURBS method expresses complex geometries using control points and associated weights as design variables\cite{liang_multi-objective_2010}. 
Its representation is inherently smooth, avoiding noise and bumps\cite{lepine_wing_2000} and facilitating local modifications. 
Nevertheless, it remains challenging to coordinate the complex relationships that must be satisfied between multiple control points for specific constraints\cite{painchaud-ouellet_airfoil_2006}, such as designing under a specified thickness. 
Similarly, Free Form Deformation (FFD) offers a wide range of generative capabilities, but its correlation with geometric features is limited\cite{yamazaki_efficient_2008}.
Although these methods can satisfy geometric constraints through optimization. 
This constraint-handling approach may lack intuitive appeal for designers. Hence, there is a critical need to develop a parametric method with high levels of intuitiveness, flexibility, and representative accuracy that can meet the demand of modern aerodynamic optimization.

\begin{figure*}[hbt!]
\label{fig1}
\centering
\includegraphics[page=1,width=.9\textwidth]{"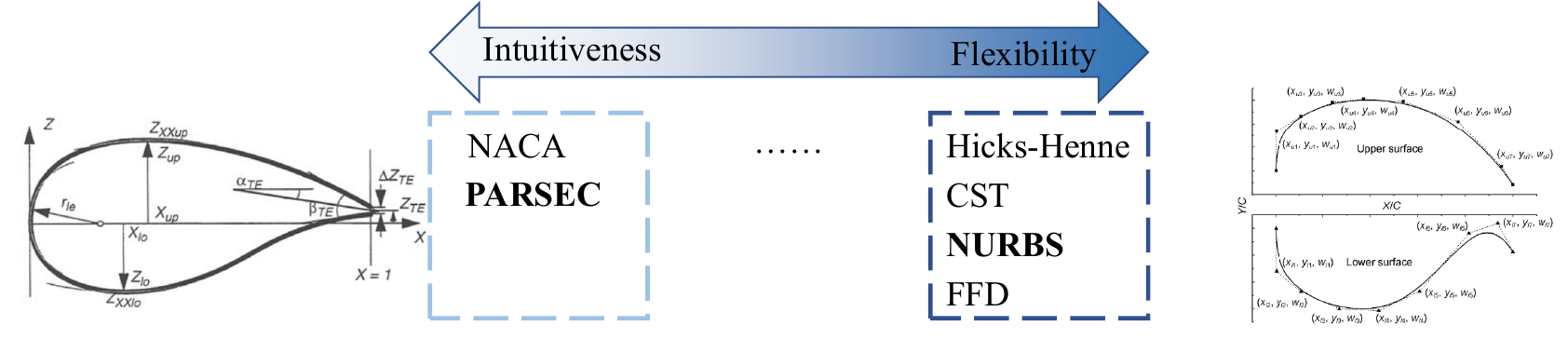"}
\caption{Classification of popular parametric approaches.}
\end{figure*}

In recent years, machine learning has revolutionized the engineering field, enabling the development of complex and comprehensive representations using data-driven generative algorithms. 
Among these, Generative Adversarial Network (GAN)\cite{goodfellow_generative_2014} has gained popularity as one of the most effective deep learning algorithms for unsupervised learning in complex distributions. 
Comprising at least two modules, the Generator generates novel data samples while the Discriminator classifies between real and fake samples. 
In contrast, Variational AutoEncoder (VAE)\cite{kingma_auto-encoding_2014} is a generative model that includes an encoder and decoder, which work towards encoding input data to a latent vector space and decoding them back to the desired output. 
VAE balances reconstruction loss and divergence from the distribution to improve the accuracy and generalization of the generative model. 
Compared to GAN, VAE is relatively easy to implement and offers greater training stability. 
Specifically, the encoder and decoder in VAE can be leveraged to fit airfoil parameters and generate new airfoil shapes, making it an attractive option for developing airfoil parametric methods. Additionally, the use of conditionally generative models, such as the Conditional Generative Adversarial Network (CGAN)\cite{mirza_conditional_2014} and Conditional Variational AutoEncoder (CVAE)\cite{sohn_learning_2015}, solves the problem of processing data under conditional constraints.

In the present research, two parametric generative schemes are proposed to represent the complicate design space effectively while satisfying specific constraints, such as features related with thickness, camber, and trailing edge. 
\textbf{1. Soft-constrained scheme}: a CVAE-based model trains geometric constraints as part of the network and can provide constrained airfoil synthesis. 
Because this scheme generate airfoils that exhibit slight deviations from the expected geometric constraints, so it is called soft-constraint scheme; 
\textbf{2. Hard-constrained scheme}: a VAE-based model serves to generate diverse airfoils and an FFD-based technique projects the generated airfoils onto the given constraints.
This scheme generate airfoils that strictly conform with constraints, although the projection may produce few odd airfoil shapes.

\section{Related works}
Principal Component Analysis (PCA) is a commonly used linear dimensionality reduction method that reduces the dimensionality of high dimensional data while retaining as much information as possible. 
The dimensionality reduction ability of PCA makes it a natural tool for parameterizing airfoils, representing the geometry of the airfoil using low dimensional modal coefficients to reduce the search space during optimization and improve optimization efficiency.
\citet{Robinson_Concise} developed a set of geometrically orthogonal base functions from a family of supercritical airfoils, which enabled a more concise representation of airfoil compared to analytical methods. 
\citet{Masters_Geometric_Comparison} presented a review of various aerofoil shape parameterization methods used in aerodynamic shape optimization, concluding that PCA-based airfoil modes have excellent performance based on their efficiency and effectiveness in terms of design variables and geometric tolerance.
\citet{li2019data} proposed a method for accurate and efficient parameterization of airfoils using camber and thickness mode shapes. Building on this work, they developed a Webfoil toolbox that allows for interactive airfoil aerodynamic optimization on any modern computing device.
However, PCA has limited representative accuracy due to the use of the linear subspace and poses a challenge for the imposition of design constraints, thereby limiting its intuitiveness.

GAN has demonstrated its ability to generate complex and diverse content from a low-dimensional latent distribution, which makes it a promising tool for airfoil synthesizing.
\citet{chen_aerodynamic_2019,chen2020airfoil} introduced BézierGAN to generate to generate Bézier curve control points by GAN, then used the control points to formulate the airfoils. The pipeline guaranteed smooth airfoil shape and reduction of feature domain. However, the Bézier curves restricted the diversity and failed to connect geometry feature explicitly. 
\citet{du_b-spline-based_2020,du_rapid_2021} proposed a similar method BSplineGAN to generate airfoils and applied it into a fast interactive aerodynamic optimization framework. 
\citet{achour_development_2020} used CGAN to generate airfoil under specific aerodynamic characteristics such as lift-to-drag ratio ($C_L/C_D$) and structural requirements (shape area). This sparked the idea of incorporating optimization objectives into deep learning models and conducting optimization through the training process. 
\citet{li_efficient_2020} proposed a deep CGAN to sample airfoils and detect the geometric abnormality quickly without using expensive computational fluid dynamic models, and embedded in a surrogate-based aerodynamic optimization framework to perform airfoil aerodynamic optimization.
While GAN has been effective in airfoil shape parameterization, it is still subject to certain limitations in direct reconstruction of the geometry.

Given that the characteristics of VAE align well with airfoil parametric methods, many researchers have explored its potential in this area. 
\citet{wang_airfoil_2021} built a VAE-GAN model combining VAE and GAN, which can encode and synthesize airfoils by interpolating/extrapolating learned features. 
\citet{wang_inverse_2021} proposed a CVAE-GAN based inverse design approach to generate airfoils for given pressure distribution, while the smoothness measurement can prove the authenticity and accuracy of airfoil. 
\citet{yonekura_inverse_2021} used CVAE to analyze the relationship between aerodynamic performance and the airfoil shape, and the airfoil can be generated for desired aerodynamics performance. However, some of the airfoils produced by this strategy may not be realistic and smooth.
Further literature on the widespread application of machine learning methods in aerodynamic shape optimization can be found in \citet{li_machine_2022}, which provides a comprehensive overview of the subject.

% constraints
Additionally, researchers have been actively exploring methods for integrating geometric constraints into parametric tools to better express design intent, limit the design space search range, and improve the efficiency of optimization design.
\citet{giammichele2007airfoil} proposed a multiresolution B-spline curve editing scheme for airfoil design and optimization that integrates thickness and camber constraints, and showed its efficacy in enhancing aerodynamic modules for multidisciplinary design optimization.
\citet{Li_Adjoint_Free,li_low-reynolds-number_2022} proposed an optimal sampling method based on deep learning that integrates a GAN model and an optimization process. By leveraging the extracted PCA modes, a more compact parameterization space is established, leading to notable enhancements in optimization efficiency for both low-Reynolds-number airfoil design and single-point/multi-point wing design.

Building upon previous research in airfoil parameterization, we seek to propose an innovative approach that utilizes deep learning methods to develop a more efficient and straightforward parameterization method for airfoils. This method is intended to facilitate both manual and automatic design optimization, while accommodating the designer's comprehension of geometric parameters and being unrestricted by specific optimization problems or design processes.

\section{Methodology}

\subsection{Dataset}
The utilization of appropriate datasets is fundamental to the efficacy and generalization of the model training and validation process. To fulfill the varying requirements of the research, three distinct datasets have been employed.
\subsubsection{UIUC Dataset (Training)}
The training dataset is obtained from the UIUC Coordinates Database at the University of Illinois Urbana-Champaign, which includes more than 1500 airfoils with diverse types that can be applied to wind turbine blades, compressor blades, aircraft wings, and other structures. Specifically, the airfoils for aircraft wings cover low-speed, medium-speed, and supercritical airfoils. This dataset, referred to as Training in this paper, has been widely utilized in various studies\cite{achour_development_2020,li_efficient_2020,wang_airfoil_2021}.

\begin{figure*}[hbt!]
\centering
\includegraphics[page=2,width=1.\textwidth]{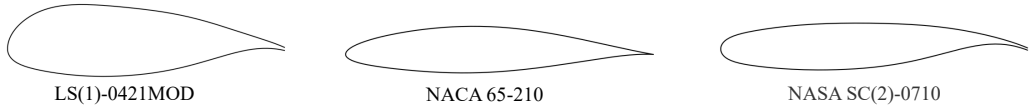}
\caption{Typical airfoils in UIUC dataset.}
\end{figure*}

\subsubsection{Random NACA 4 Digits Dataset (Test1)}

The NACA Four-digits airfoil is a well-known and straightforward method for parameterizing airfoils, which uses only three parameters to describe an airfoil: the maximum camber, the relative position of the maximum camber, and the maximum thickness. An example is NACA 2412, which indicates a camber of 2\% of the chord, a position of maximum camber at 40\% of the chord, and a maximum thickness of 12\% of the chord. To generate the Random NACA 4 digits Dataset, 1500 airfoils are randomly generated using more refined values based on this definition. Latin Hypercube Sampling (LHS) is used to sample the three parameters within the range of existing NACA airfoils in the UIUC dataset. Since the UIUC dataset includes many NACA airfoils, the Random NACA 4 digits Dataset can be viewed as an interpolation of the UIUC dataset and will serve as the "interpolation" test set (denoted by Test1 in this paper).

\begin{figure*}[hbt!]
\centering
\includegraphics[page=3,width=1.\textwidth]{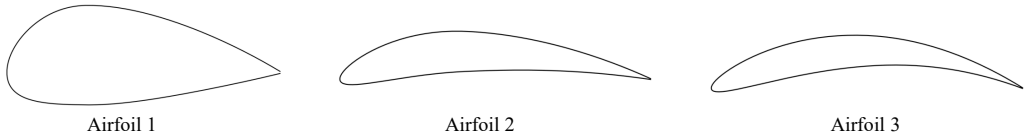}
\caption{Typical airfoils in random NACA 4 digits dataset.}
\end{figure*}

\subsubsection{Supercritical Airfoil Dataset (Test2)}

The previous study\cite{wang_inverse_2021} involved the creation of a supercritical airfoil dataset from a novel airfoil design. 
The dataset exhibits diversity in both geometric features and aerodynamic performance and includes constraints such as the leading edge radius being greater than or equal to 0.007 and the drag coefficient not exceeding 0.1, in order to avoid impractical airfoils. 
Since none of the airfoils in this dataset are included in the UIUC dataset, it serves as an \textbf{extrapolation} test set (denoted by Test2 in this paper).

\begin{figure*}[hbt!]
\centering
\includegraphics[page=4,width=1.\textwidth]{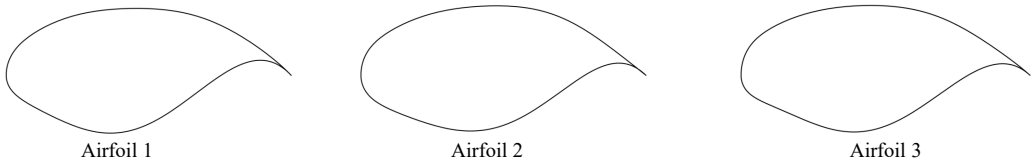}
\caption{Typical airfoils in supercritical airfoil dataset.}
\end{figure*}

\begin{figure}[hbt!]
\centering
\includegraphics[page=5,width=.48\textwidth]{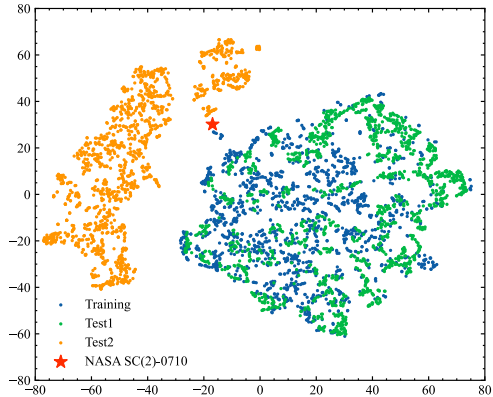} 
\caption{t-SNE plot of all datasets.}
\label{fig5}
\end{figure}

To visualize the distribution of all the samples involved in our study, the t-Distributed Stochastic Neighbor Embedding (t-SNE) technique is employed to reduce the dimensionality of each sample. Fig.~\ref{fig5} presents the t-SNE plot of all three datasets, which clearly shows the interrelationships and boundaries among them. 
The Training dataset is widely distributed, while the Test1 dataset overlaps with Training. 
In contrast, Test2 is significantly different from the other two datasets, and the closest sample in Training to Test2 is NASA SC(2)-0710, which is a member of a family of supercritical airfoils with varying thicknesses and design lift coefficients\cite{harris_nasa_1990}.

\subsection{Pre-processing}

The pre-processing of the dataset includes several steps. Firstly, a rotation/scale transformation is applied to transform the leading edge and trailing edge points of each airfoil to [0,0] and [1,0] in Cartesian coordinates. Secondly, resampling is used to create a homogeneous representation of all airfoils, since they have different coordinate formats due to diverse sources. 
The cosine distribution of x-coordinates is given by Eq.~\eqref{1:equation}:

\begin{equation}
\label{1:equation}
\begin{array}{l}
\theta_{i}=\frac{\pi(i-1)}{N} \\
x_{i}=1-\cos \left(\theta_{i}\right)
\end{array}
\end{equation}
where $N$ is set to 201 to ensure 100 segments on both upper and lower surfaces. A cubic spline curve is used to interpolate the corresponding y coordinates, allowing the airfoil geometry to be represented solely by the y coordinates. 
The third step involves feature extraction, where key geometry features such as maximum thickness, maximum camber, and leading edge radius are extracted. In the final step, normalization is applied independently to both y coordinates and geometry features. The y coordinates are normalized based on coefficients derived from the training set, which scale the training set to a range of [-1,1]. The normalization of geometry features differs slightly, with each feature normalized independently using its corresponding coefficients to a range of [-1,1].

\begin{figure*}[hbt!]
\centering
\subfigure[Geometry feature distribution.]{
\includegraphics[page=6,width=.45\textwidth]{"figs.pdf"}}
\subfigure[Objective space distribution.]{
\includegraphics[width=.45\textwidth]{"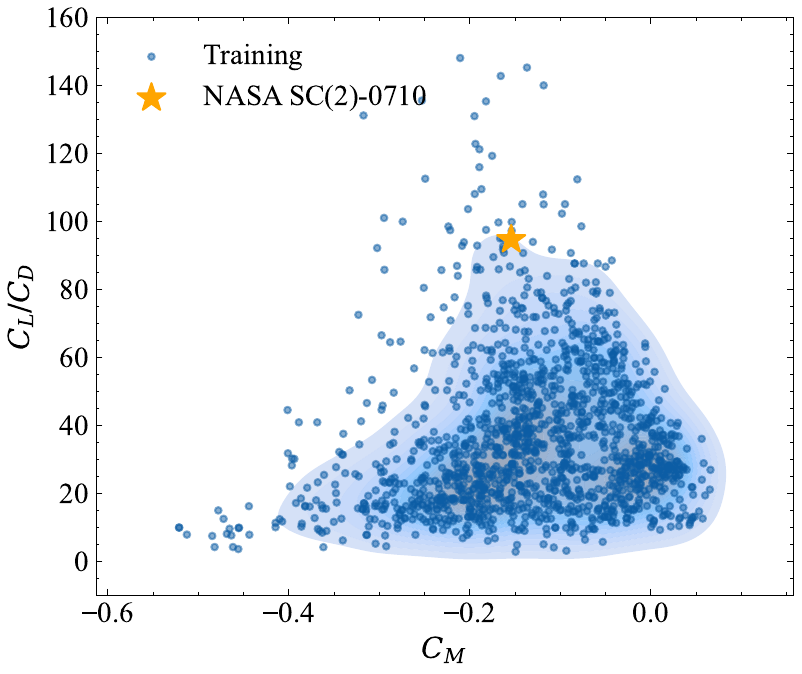"}}
\caption{Feature distribution of datasets.}
\label{fig6}
\end{figure*}

The diversity of the geometric features of the dataset is visualized in Fig.~\ref{fig6}a, which depicts the distribution of the maximum camber and maximum thickness in the dataset. 
A broad and uniform pattern is observed in the Training set, with the exception of a significant number of zero camber samples. Conversely, the Test1 set displays a uniform distribution within the prescribed rectangular range. However, the feature distribution in the Test2 set is highly concentrated and characterized by an identical maximum thickness value.
In addition, the diversity of the objective parameters space is also visualized. 
To evaluate the objective space distribution, an in-house code based on the corrected full potential equation is utilized for quick aerodynamic performance evaluation. 
The scatter and kernel density estimate (KDE) plot in Fig.~\ref{fig6}b shows the relationship between the lift-to-drag ratio and moment coefficient ($C_M$) of the Training set, calculated at a constant lift coefficient of 0.70 and a Mach number of 0.77, which is suitable for supercritical airfoils. 
The plot indicates that the NASA SC(2)-0710 airfoil performs well in terms of lift-to-drag ratio and is located close to the 0.1 threshold of the KDE plot.

\subsection{Models}
\subsubsection{VAE}

\begin{figure*}[hbt!]
\centering
\includegraphics[page=7,width=.9\textwidth]{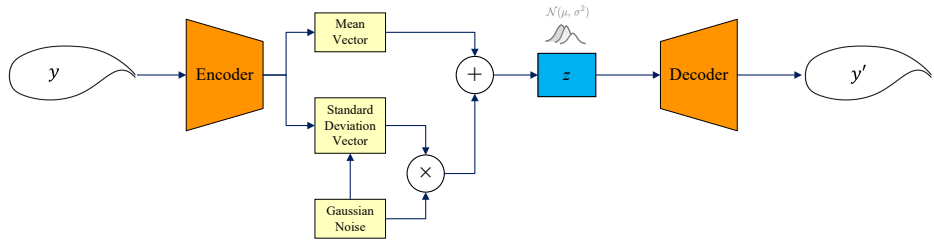}
\caption{The base VAE architecture.}
\label{fig7}
\end{figure*}

The base VAE architecture shown in Fig.~\ref{fig7} is the foundation of generative schemes. The airfoil coordinates $y$ is taken as the input, and the reconstructed airfoil coordinates is denoted as $y^{\prime}$. The encoder and decoder are constructed by MLP with similar structure.
The encoder maps $y$ to learned latent distribution z, and the decoder generate $y^{\prime}$ from the latent vector $z$. The encoder and decoder can be expressed as:
\begin{equation}
z \sim \operatorname{Enc}(y)=q(z \mid y), \quad y^{\prime} \sim \operatorname{Dec}(z)=p\left(y^{\prime} \mid z\right)
\label{2:equation}
\end{equation}
where $q(z)$ is the learned distribution and $p(z)$ is the prior distribution. The loss function of VAE is given by:
\begin{equation}
L_{VAE}=\alpha_{1} L_{mse}+\alpha_{2} L_{kld}
\label{3:equation}
\end{equation}
where $L_{mse}=\left\|y-y^{\prime}\right\|_{2}^{2}$ is the mean squared error between the original and synthesized airfoil, and $L_{kld}=D_{kl}\left(q\left(z|y\right)||p(z)\right)$ is the Kullback-Leibler divergence between the learned and prior distribution. Minimizing $L_{mse}$ leads to higher reconstruction accuracy, while minimizing $L_{kld}$ results in more effective airfoil generation based on prior distribution. 
The generative capability can introduce errors that affect reconstruction accuracy during training, while improving reconstruction accuracy can weaken the generative capability. The hyperparameters $\alpha_1$ and $\alpha_2$ are used to balance these factors.

\subsubsection{Soft-constrained scheme}

\begin{figure*}[hbt!]
\centering
\includegraphics[page=8,width=.9\textwidth]{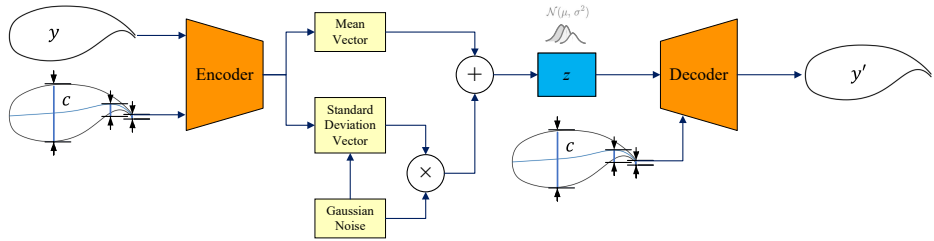}
\caption{The soft-constrained scheme.}
\label{fig8}
\end{figure*}

The soft-constrained scheme is developed based on CVAE. In CVAE, the Encoder maps $y$ to $z$ with distribution $q\left(z \middle| y,c\right)$ under specific given geometric constraints $c$. The Decoder generates $y^\prime$ with latent vector $z$ under constraints $c$. The loss function is the same as VAE. 
It is obvious seen that the geometry constraint $c$ exactly matches the original sample $y$, but there is a certain error with the generated airfoil $y^\prime$, which can be only reduced but not eliminated with the training process.

\subsubsection{Hard-constrained scheme}

\begin{figure*}[hbt!]
\centering
\includegraphics[page=9,width=.9\textwidth]{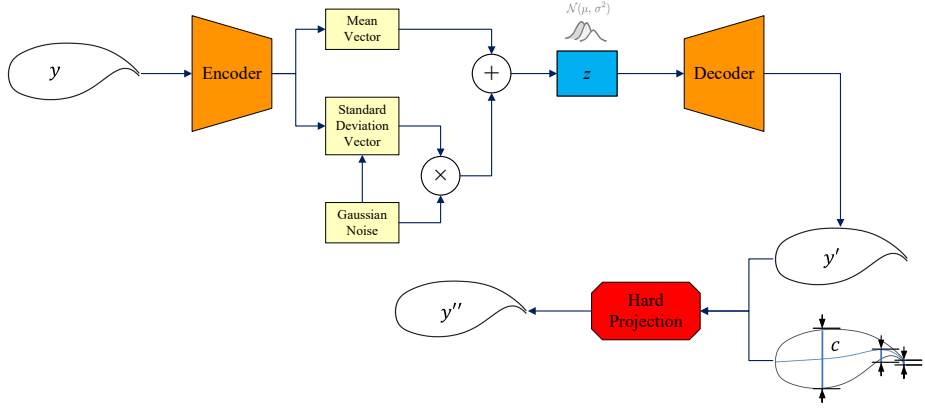}
\caption{The hard-constrained scheme.}
\label{fig9}
\end{figure*}

The hard-constrained scheme combines VAE and a hard projection module, in which VAE is used to generate diverse airfoils $y'$ that may not satisfy the required geometric constraints $c$. 
The hard projection module establishes a one-to-one mapping from $y'$ to $y''$, ensuring that $y''$ strictly satisfies the geometric constraints $c$.
$L_{mse}$ for hard-constrained scheme measures the difference between $y$ and $y''$.

The hard projection is achieved through the FFD parameterization tool, which involves mapping the geometric parameters on the canvas corresponding to the constraints, followed by stretching and modifying the canvas using FFD to satisfy the corresponding constraints. 
The 2D Bezier-based FFD is defined in terms of a bivariate Bernstein polynomial. The displacement $\Delta x(s,t)$ of any point $x(s,t)$ in the control box can be expressed as:
\begin{equation}
x(s, t)+\Delta x(s, t)=\sum_{i}^{l} \sum_{j}^{m}\left[B_{l-1}^{i-1}(s) B_{m-1}^{j-1}(t)\right]\left[P_{i, j}+\Delta P_{i, j}\right]
\label{4:equation}
\end{equation}
where $P_{i,j}$ and  $\Delta P_{i, j}$ are the original coordinates and displacements of the vertices of the control box, respectively. The Eq.~\eqref{4:equation} can be rewritten into the matrix form: 
\begin{equation}
\Delta \mathbf{x}=\mathbf{B} \cdot \Delta \mathbf{P}
\label{5:equation}
\end{equation}

In the concurrent application, $\Delta\mathbf{x}$ is obtained from the geometry constraints, $\mathbf{B}$ remains constant when building up the control box, and $\Delta\mathbf{P}$ is the only unknown variable.
In this particular application, $\Delta\mathbf{x}$ is determined by the difference between the geometric feature of $y'$ and the corresponding geometry constraints. 
$\mathbf{B}$ remains constant once building up the control box, and $\Delta\mathbf{P}$ is the only unknown variable.

\begin{figure}[hbt!]
\centering
\includegraphics[page=10,width=.45\textwidth]{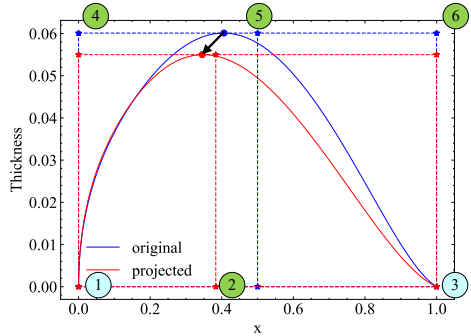}
\caption{Projection of thickness distribution.}
\label{fig10}
\end{figure}

% Fig.~\ref{fig10} illustrates a simple example to project the thickness distribution to the given constraints. 
Fig.~\ref{fig10} depicts a simple example of projecting the thickness distribution to the specified constraints. 
Initially, the airfoil shape is converted into a representation of camber and thickness, after which constraints are imposed based on the thickness distribution.
The given constraints are the maximum thickness and maximum thickness position, which can be expressed as the peak point of the thickness distribution curve.
The main purpose of FFD-based hard projection is to move the peak point to the location of the specified constraint in a unique manner while maintaining the smoothness of the thickness distribution.
Thus, $\Delta \mathbf{x}$ includes 2 variables([$dx,dy$] of peak point).
To ensure full coverage of the thickness distribution while avoiding the influence on high-order derivatives of the curve, a control box comprising of 6 control vertices is selected. $\Delta \mathbf{P}$ encompasses 12 unknowns([$dx,dy$] of 6 control vertices).
Eq.(5) is currently underdetermined and has an infinite set of solutions. To obtain a unique solution with improved curve characteristics, the degree of freedom of the control box is partially released.
In this case, vertex 2/5 share the same $dx$, vertex 4/5/6 share the same $dy$, while other degrees of $\Delta \mathbf{P}$ are frozen.
Then $\Delta \mathbf{P}$ can be solved for 2 unknowns, and deformation of any other points on the curve can be derived from the equation.
Other constraints can be satisfied in the similar way.

The combination of VAE and FFD projection enables the hard-constrained scheme to have generative capabilities while strictly adhering to geometry constraints.

\section{Numerical experiment}

\subsection{Hyperparameters}

The hyperparameters of the generative models are tuned through the VAE model, just relying exclusively on the training dataset. Therefore, the model is unfamiliar with the Test1 and Test2 sets.
The encoding network comprises three hidden layers containing 256, 128, and 64 neurons, while the decoding network has three hidden layers containing 64, 128, and 256 neurons. The latent vector $z$ is initially set to a dimension of 10 based on previous research on flow field reconstruction\cite{wang2021flow}.
The design is governed by five geometry constraints, namely the maximum thickness position, maximum thickness, trailing edge thickness, maximum camber position, and maximum camber, which serve as the condition $c$.
The activation function utilized in the present study is the hyperbolic tangent (Tanh), which is applied to all layers of the encoder and decoder except for the output layer. The Adam optimizer is employed to minimize the loss function. The training process comprises 5000 epochs with a learning rate initialized at 5.0e-4 and decays to 5.0e-5 after 2500 epochs.

Furthermore, the present study places emphasis on the critical hyperparameters, including the type of prior distribution and $\alpha_1$ value in the loss function, as they can cause the collapse or divergence of the latent space vectors. It is worth noting that the choice of prior distribution can affect the performance of the VAE model. Specifically, the VAE model that employs Gaussian variational prior distribution (referred to as N-VAE) often experiences collapse in low dimensions and 'soap-bubble-effect' in high dimensions. To address these issues, the researchers have developed a VAE model with a hyperspherical latent space\cite{davidson_hyperspherical_2018}, referred to as S-VAE. Besides, adjusting the value of $\alpha_1$ can also prevent the occurrence of collapse.

% In addition, two crucial hyperparameters, prior distribution type and $\alpha_1$ in loss function, will be discussed in detail. These two factors will be discussed together because both of them can lead to collapse or divergence of the latent space vectors. VAE using the Gaussian variational prior distribution (denoted by N-VAE in this section), shows collapse in low dimensions and ‘soap-bubble-effect’ in high dimensions. To solve these issues, VAE with a hyperspherical latent space\cite{davidson_hyperspherical_2018} (denoted by S-VAE in this section) is developed. Besides, $\alpha_1$ can also help prevent collapse.

\begin{figure*}[hbt!]
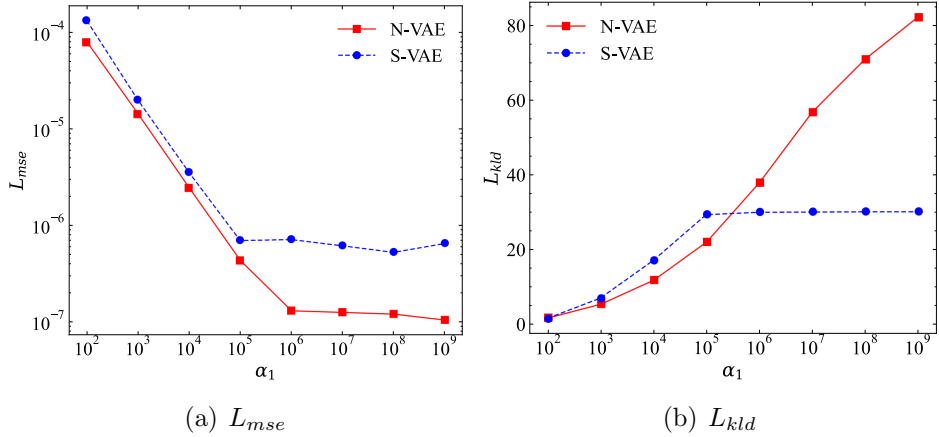

  \centering
  \subfigure[$L_{mse}$]{\includegraphics[page=11,width=.45\textwidth]{"figs.pdf"}}
  \subfigure[$L_{kld}$]{\includegraphics[page=12,width=.45\textwidth]{"figs.pdf"}}
  \caption{Influence of distribution type and $\alpha_1$.}
  \label{fig11}
\end{figure*}

Fig.~\ref{fig11} illustrates the influence of distribution type and $\alpha_1$ on the loss function of VAE model, indicating the competitive relationship between \ $L_{mse}$ and $L_{kld}$. As $\alpha_1$ increases from ${10}^2$ to ${10}^6$, $L_{mse}$ of both N-VAE and S-VAE decreases and remains almost constant after $\alpha_1$ reaching ${10}^6$. $L_{mse}$ of the N-VAE is smaller than S-VAE for all $\alpha_1$, indicating higher reconstruction accuracy of N-VAE. On the other side, $L_{kld}$ of the N-VAE keeps increasing as $\alpha_1$ increases while $L_{kld}$ of the S-VAE remains unchanged after ${10}^5$. If $L_{kld}$ is greater than 30.0, there are a lot of bad airfoils when sampling from the prior distribution. Therefore, S-VAE succeeds in preventing the divergence of $L_{kld}$. Overall, N-VAE performs better in reconstruction accuracy, and SVAE is able to maintain strong generative ability even when $\alpha_1$ is quite large. When comparing the distribution type and $\alpha_1$, N-VAE at $\alpha_1={10}^5$ appears to be the best choice to balance the accuracy and generative capability.

\begin{figure*}[hbt!]
  \centering
  \subfigure[$L_{mse}$]{\includegraphics[width=.45\textwidth]{"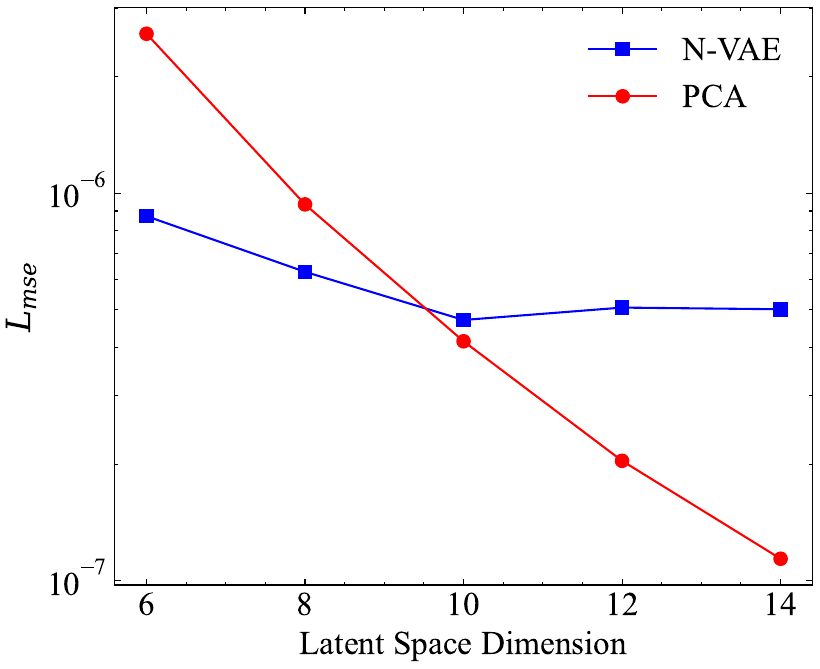"}}
  \subfigure[$L_{kld}$]{\includegraphics[width=.45\textwidth]{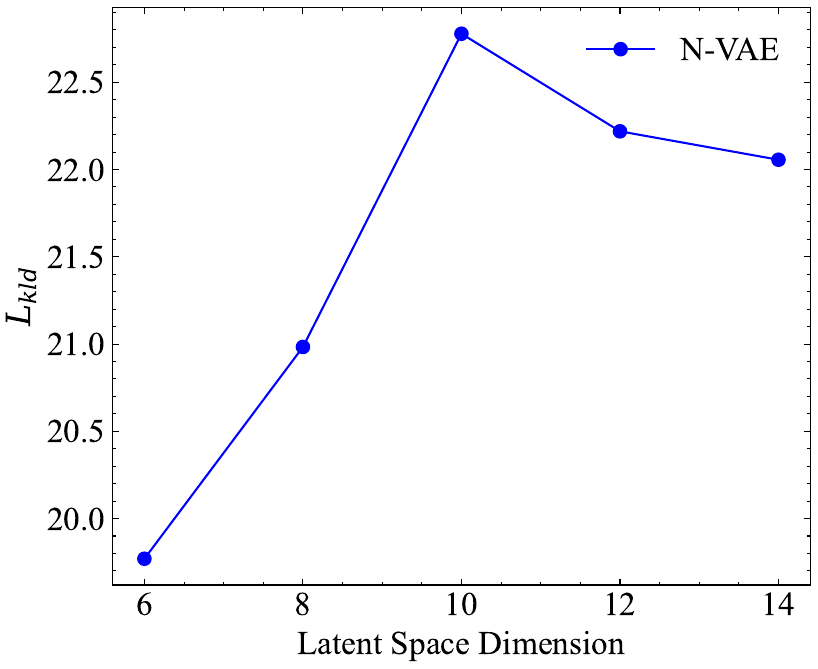}}
  \caption{Influence of latent space dimension.}
  \label{fig11+}
\end{figure*}

The influence of the latent space dimension on the results is further investigated, with the airfoil linear mode decomposition based on PCA taken as a benchmark for reconstruction error $L_{mse}$, as shown in Fig.~\ref{fig11+}. 
The reconstruction error of PCA displays a rapid decrease with increasing number of modes. 
In contrast, the reconstruction error of VAE declines as the dimension of the latent space increases when it is below 10, but it slightly increases when the dimension is greater than 10.
The reconstruction errors of both methods exhibit similar performance when the dimension equals 10.
On the other hand, $L_{kld}$ increases as the dimension of the latent space increases when it is less than 10, but decreases gradually beyond this threshold.
Notably, the choice of $\alpha_1$ plays a crucial role in the analysis of the latent space dimension. 
Based on our results, setting $\alpha_1$ to 1e-5 yields a preferable latent space dimension of 10.

The VAE serves as a fundamental component for both the soft and hard schemes, and thus, minor adjustments to the neural network have minimal effect on the selection of hyperparameters. Both schemes utilize identical hyperparameters and exhibit satisfactory performance.

\subsection{Airfoils Reconstruction}

\begin{table}[hbt!]
\caption{\label{tab:table1} Results of reconstruction errors}
\centering
\begin{tabular}{lcccc}
\toprule
% \hline
& \multicolumn{2}{c}{Soft-constrained}&  \multicolumn{2}{c}{hard-constrained}\\\cline{2-5}
Datasets& MSE & MAE& MSE& MAE \\\midrule
Training & 3.60E-07 & 4.51E-04 & 5.36E-07 & 5.46E-04 \\ 
Test1 & 5.53E-07 & 5.22E-04 & 8.42E-07 & 6.35E-04 \\ 
Test2 & 2.50E-06 & 1.14E-03 & 2.13E-06 & 1.11E-03 \\ 
\bottomrule
\end{tabular}
\end{table}

% \begin{table}[<options>]
% \begin{tabular*}{\tblwidth}{@{}LL@{}}
% \toprule
% \midrule
%  & \\
%  & \\
%  & \\
%  & \\
% \bottomrule
% \end{tabular*}
% \end{table}

\begin{figure*}[hbt!]
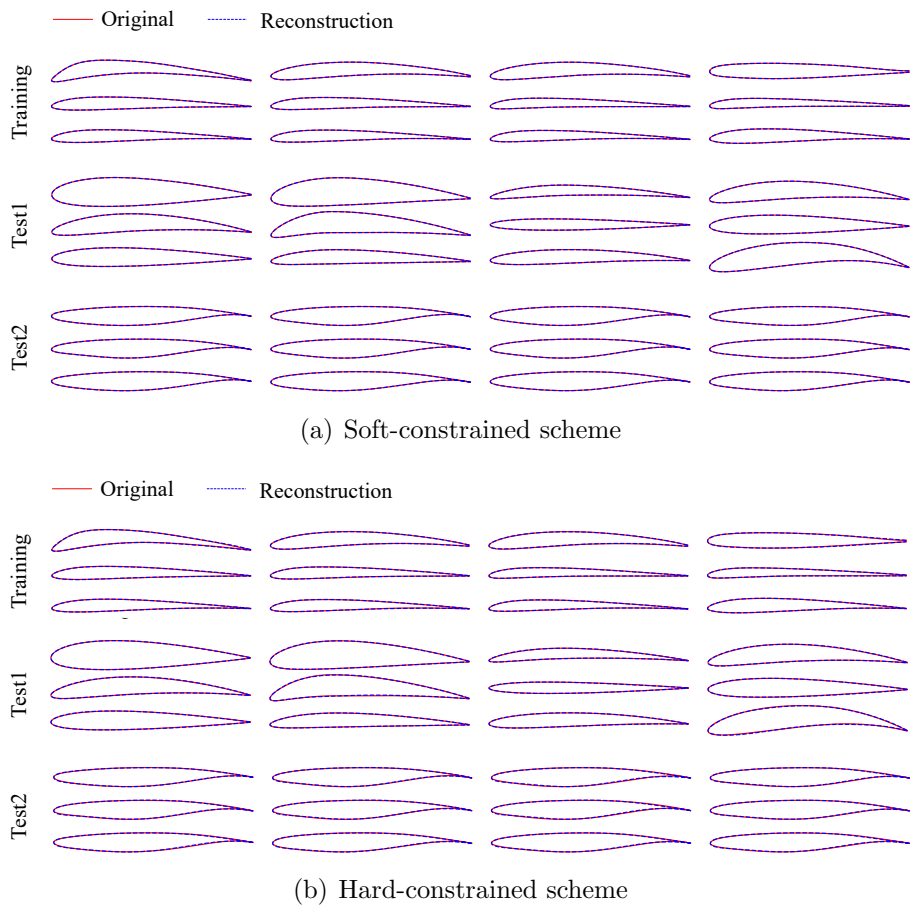

  \centering
  \subfigure[Soft-constrained scheme]{\includegraphics[page=13,width=.9\textwidth]{"figs.pdf"}}
  \subfigure[Hard-constrained scheme]{\includegraphics[page=14,width=.9\textwidth]{"figs.pdf"}}
  \caption{Comparisons of reconstructed airfoils.}
  \label{fig12}
\end{figure*}

Table.~\ref{tab:table1} presents the errors for the two generative schemes. It is observed that the soft-constrained scheme achieves smaller mean squared error (MSE) and mean absolute error (MAE) on both the Training and Test1 set, whereas the hard-constrained scheme performs slightly better on the Test2 set. The error values for Test2 are higher than those for Test1, indicating that extrapolation is more challenging than interpolation. Fig.~\ref{fig12} shows the randomly selected original airfoils alongside their corresponding reconstructed airfoils, which are all smooth and exhibit good agreement with the original airfoils.

\begin{figure*}[hbt!]
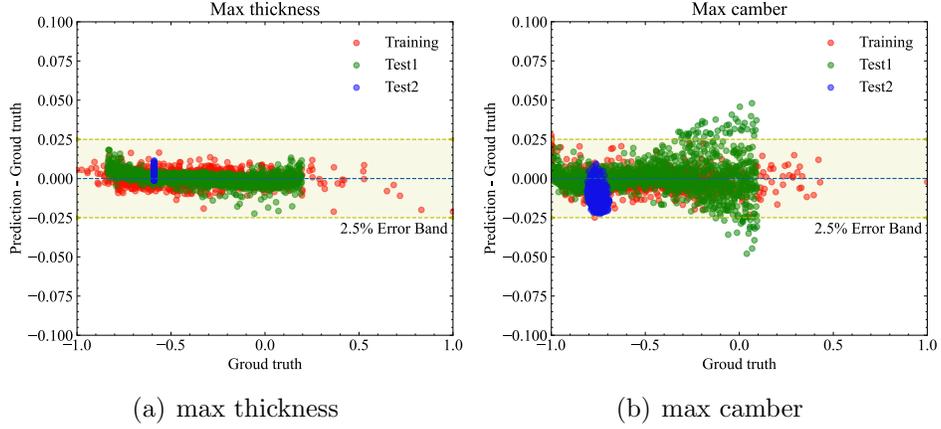

  \centering
  \subfigure[max thickness]{\includegraphics[page=15,width=.45\textwidth]{"figs.pdf"}}
  \subfigure[max camber]{\includegraphics[page=16,width=.45\textwidth]{"figs.pdf"}}
  \caption{Error distribution of geometry features for soft-constrained scheme.}
  \label{fig13}
\end{figure*}

The hard-constrained scheme produces reconstructed airfoils with geometric features that are exactly the same as the original airfoils due to the precise projection. However, some errors are present in the soft-constrained scheme. The error distributions for the maximum thickness and maximum camber are shown in Fig.~\ref{fig13}, indicating that a 2.5\% error band covers all samples in all datasets for the maximum thickness, while 98.5\% of the samples are covered by the error band for the maximum camber.

\subsection{Airfoils Synthesizing}
% In generative schemes, airfoils are encoded into latent space first, then the latent vectors are employed as variables to synthesize new airfoils, with no explicit relationship between the geometric characteristics and the latent space. The latent vector distribution of VAE, for example, is shown in Fig.~\ref{fig14}. For each dimension, the Training set has been well-trained, with Test1 and Test2 clustered in specific regions. In particular, $z5$ of Test2 is distributed in a very narrow band, looking back to Fig.~\ref{fig5}, the thickness distribution of Test 2 is also in a thin band, this could be a possible mapping. It demonstrates that it is possible to control the geometric features to some extent according to the characteristics of latent vector distribution.

\begin{figure*}[hbt!]
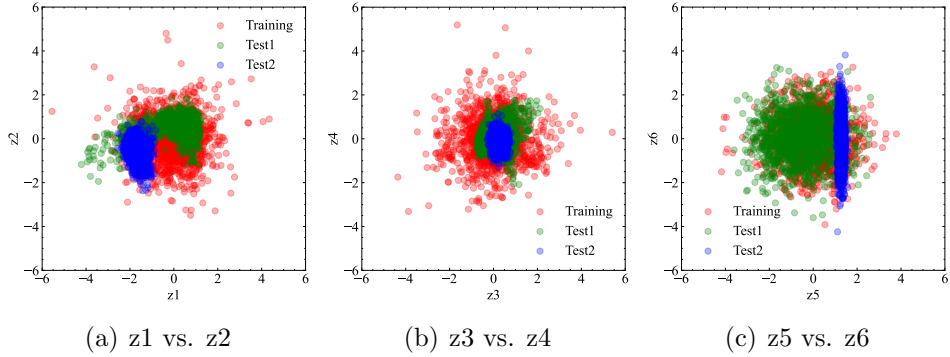

  \centering
  \subfigure[z1 vs. z2]{\includegraphics[page=17,width=.3\textwidth]{"figs.pdf"}}
  \subfigure[z3 vs. z4]{\includegraphics[page=18,width=.3\textwidth]{"figs.pdf"}}
  \subfigure[z5 vs. z6]{\includegraphics[page=19,width=.3\textwidth]{"figs.pdf"}}
  \caption{Latent vector distribution of VAE.}
  \label{fig14}
\end{figure*}

In the two generative schemes, airfoil geometries are encoded into a latent space, which usually has little explicit link with geometric characteristics. Fig.~\ref{fig14} presents the distribution of the latent vectors for VAE, where the Training set is well-trained, and Test1 and Test2 are clustered in specific regions for each dimension. Notably, $z5$ of Test2 is distributed within a very narrow band, which is similar to the thickness distribution of Test2 as demonstrated in Fig.~\ref{fig5}
This finding suggests a possible correlation between the geometric features and the latent space. Therefore, it can be inferred that the proposed schemes have potential to control the geometric features to some extent based on the characteristics of the latent vector distribution.

% Airfoil synthesizing based on the latent vectors is shown in Fig.~\ref{fig15}. The reference airfoil is generated from a reference latent vector $\mathbf{z}_{\mathbf{ref}}=\mathbf{0}\in\mathbb{R}^{10}$. The synthesizing airfoils are generated with $z1$ to $z6$ varing from -4.0 to 4.0 and other vectors equaling the reference vector. It is seen that the deformation caused by $z1$ is mainly located near the trailing edge, while $z2$ affects both the leading and trailing edges. $z4$ primarily controls the max thickness position. $z3$ seems to play an important role in smoothness, which is a higher order representation of airfoil. $z5$ is clearly related to the max thickness, whereas $z6$ is critical for adjusting the max camber position. The preceding comparisons show that different latent vectors affect airfoil geometry at various scales and in different areas at various orders. The variation space of $z$ covers almost all samples, representing that the range of the geometric features variation of the synthesized airfoils can match the whole datasets, achieving the completeness of geometry space. 

\begin{figure*}[hbt!]
\centering
\includegraphics[page=20,width=1.\textwidth]{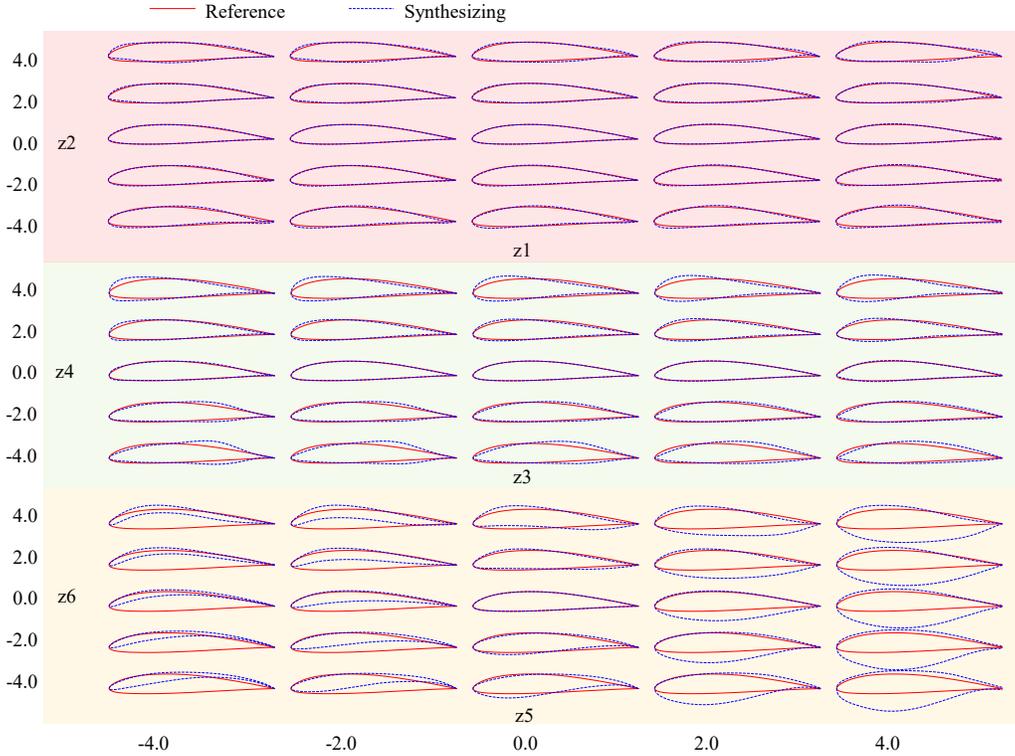}
\caption{Airfoil synthesizing of VAE.}
\label{fig15}
\end{figure*}

The results of airfoil synthesis based on the latent space is illustrated in Fig.~\ref{fig15}. The reference airfoil is generated with the latent vector $\mathbf{z}_{\mathbf{ref}}=\mathbf{0}\in\mathbb{R}^{10}$, and the synthesizing airfoils are generated with $z1$ to $z6$ varying from -4.0 to 4.0 while the other vectors remain equal to the reference vector. The results indicate that different latent vectors affect the airfoil geometry at different scales and in different areas at various orders. Specifically, the deformation caused by $z1$ is mainly located near the trailing edge, while $z2$ affects both the leading and trailing edges. $z4$ primarily controls the maximum thickness position, and $z3$ appears to play an important role in smoothness, which is a higher-order representation of the airfoil. $z5$ is evidently related to the maximum thickness, whereas $z6$ is critical for adjusting the maximum camber position. The comparison of different latent vectors highlights the completeness of the geometry space, as the variation space of $z$ covers almost all samples, allowing for the synthesis of airfoils with varying geometric features.

% For optimization design based on existing airfoil geometries, soft and hard-constrained schemes can be employed as a sampling tool to provide a set of geometric parametric variables that allow for the generation of numerous novel airfoils while still adhering to the specified constraints.

\begin{figure*}[hbt!]
  \centering
  \subfigure[Soft-constrained scheme]{\includegraphics[page=21,width=.45\textwidth]{"figs.pdf"}}
  \subfigure[Hard-constrained scheme]{\includegraphics[page=22,width=.45\textwidth]{"figs.pdf"}}
  \subfigure[PCA]{\includegraphics[width=.45\textwidth]{"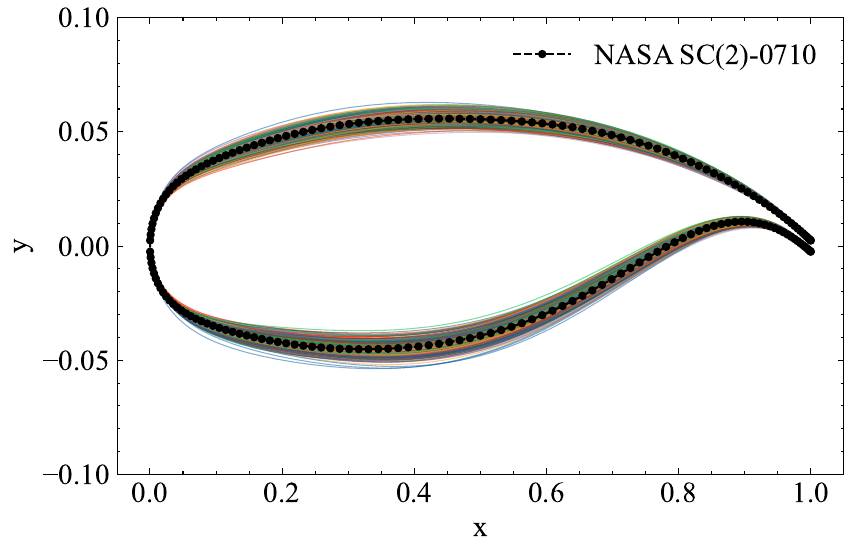"}}
  \subfigure[CST]{\includegraphics[width=.45\textwidth]{"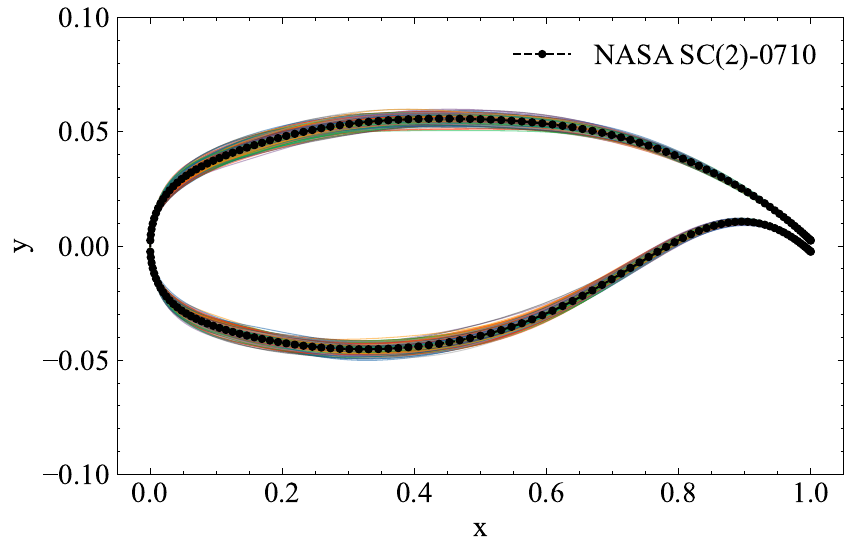"}}
  \caption{Airfoil synthesizing by different schemes.}
  \label{fig16}
\end{figure*}

Fig.~\ref{fig16} presents the airfoil synthesizing using different schemes, including the soft-constrained scheme, hard-constrained scheme, PCA, and CST. 
Each scheme generates 500 airfoils based on NASA SC(2)-0710. 
The proposed constrained schemes synthesize new airfoils by applying random variations in the range [-1, 1] using LHS on the latent vector of the reference airfoil. 
As for PCA, the modal coefficients are treated as variables, and LHS is conducted within a range of 0.01 times the distribution range of the modal coefficients in the Training set. 
The sampled coefficients are then added to the modal coefficients of the reference airfoil to generate new airfoils. 
For CST, each coefficient is subjected to LHS within a range of 0.01 times the distribution range of the coefficient in the Training set, and added to the CST coefficient of the reference airfoil to synthesize new airfoils.
It is observed that each of the methods is capable of generating airfoils with a relatively uniform distribution in the geometric space, indicating their ability to produce variations around the reference airfoil. 
It is observed that the hard-constrained scheme has a greater variance than the soft-constrained scheme, indicating that it covers a wider range of geometric space. 
Besides, the airfoils generated by the soft-constrained scheme are all smooth and reasonable, whereas some of the airfoils generated by the hard-constrained scheme are too anomalous to be considered.
The occurrence of such anomalous airfoils can be attributed to FFD application to large deformations. 
VAE generates airfoils with smaller maximum camber, but the enforced constraints of FFD map them to a larger extent, resulting in substantial distortion of the FFD canvas.
However, the proportion of these abnormal airfoil shapes is very low, with only about 10 out of 500 airfoils appearing noticeably abnormal. 
Furthermore, these abnormal airfoil shapes can be easily eliminated by controlling the range of perturbation of the variables.
Regarding the PCA and CST methods, the generated airfoils are well-distributed around the reference airfoil, and their distribution falls within a reasonable range. However, most of the new airfoils generated by these two methods do not satisfy geometric constraints.

\begin{figure*}[hbt!]
\centering
\subfigure[Soft-constrained scheme]
{\includegraphics[width=0.45\textwidth]{"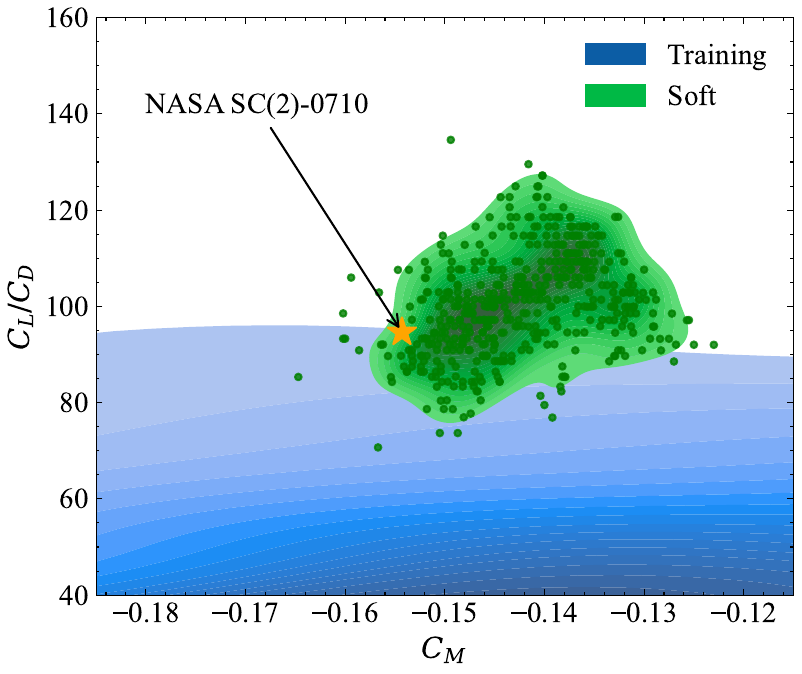"}}
\subfigure[Hard-constrained scheme]
{\includegraphics[width=0.45\textwidth]{"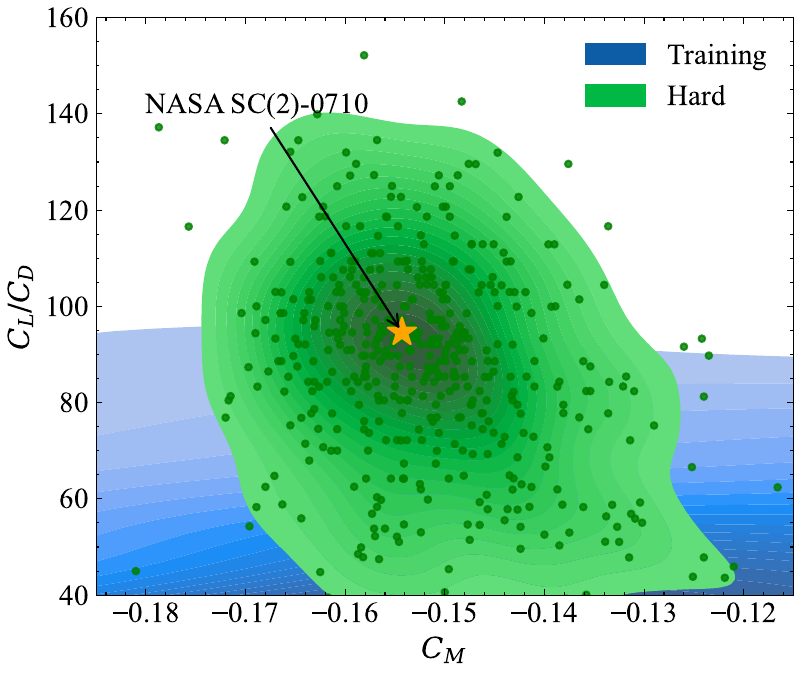"}}
\subfigure[PCA]
{\includegraphics[width=0.45\textwidth]{"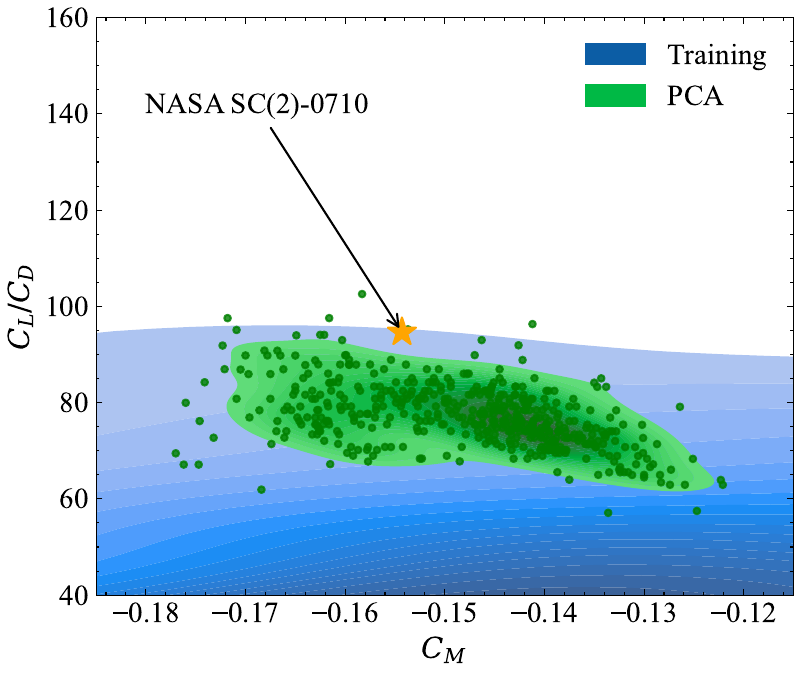"}}
\subfigure[CST]
{\includegraphics[width=0.45\textwidth]{"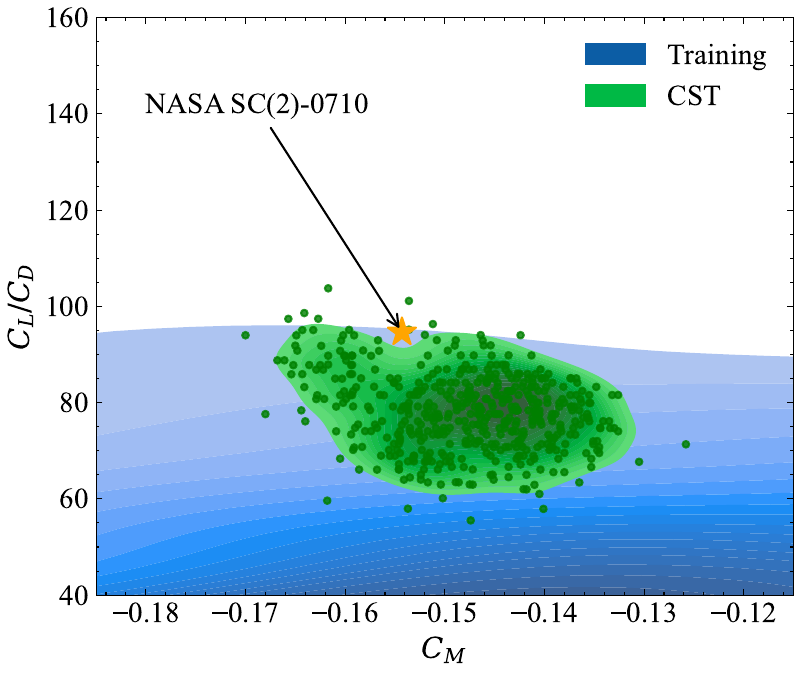"}}
\caption{Objective space distribution.}
\label{fig17}
\end{figure*}

In addition to the design space, the diversity of the objective space is also concerned.
Fig.~\ref{fig17} illustrates the distribution of lift-to-drag ratio $C_L/C_D$ and moment coefficient $C_M$ for airfoils synthesized using different schemes in both scatter and corresponding KDE plot. 
Upon a detailed comparison of two proposed methods, it is evident that the airfoil samples generated using the soft-constrained scheme demonstrate tighter clustering around the reference point, despite some degree of distribution bias. 
Furthermore, the newly generated airfoils exhibit a greater proportion of samples with $C_L/C_D$ exceeding that of the reference point, in addition to a significantly higher $C_M$ than the reference point. 
Conversely, the hard-constrained scheme yields a more diverse range of airfoil samples with an isotropic distribution, centered around the reference point, without notable bias. 
The newly generated samples consist of those with considerably high or low $C_L/C_D$; however, the distribution of performance metrics concerning the reference point is relatively balanced.
In the absence of geometric constraints, PCA and CST yield a limited number of new samples with $C_L/C_D$ exceeding the reference point, with the overall sampling data remaining within the training set boundary. 
However, the reasons for the inability of these methods to surpass the boundary may differ.
The PCA method relies on the dataset, as exemplified by the tailored airfoil modes proposed in \cite{li_low-reynolds-number_2022}. 
When the dataset satisfies certain geometric constraints, the newly generated samples tend to satisfy these constraints.
In this study, most of the airfoils in the Training set have a relatively low $C_L/C_D$ under the supercritical condition, which makes it difficult for the sampled airfoils to achieve a higher $C_L/C_D$.
The CST method characterizes the airfoil using coefficients that have a certain coupling relationship, and randomly perturbing these coefficients can result in a large number of aerodynamically infeasible geometries, reducing the $C_L/C_D$ performance.

In summary, the two schemes proposed in this study exhibit better properties than PCA and CST and can be utilized in different scenarios based on their unique distribution characteristics in the objective space. 
The soft-constrained scheme, despite its potential biased distribution resulting in objective space blind spots, has higher sampling efficiency and effectiveness, as well as a concentrated distribution near the reference airfoil. 
Thus, it is more suitable for local optimization of a high-performance airfoil, which can be initiated after finding a local optimum. 
In contrast, the hard-constrained scheme yields a significantly wider distribution range with isotropic uniformity surrounding the reference point, indicating no significant bias. 
It is more appropriate for initial sampling in global optimization to explore better extreme performance. 
However, a significant number of low-performing samples may result in decreased optimization efficiency.

\section{Conclusions}
The present research develops a novel method for parameterizing airfoils based on data-driven approaches.
VAE based generative schemes with soft and hard-constraints are proposed, which permit the encoding of arbitrary airfoil geometries and the synthesizing of airfoil that satisfy the given constraints. 
The primary findings are summarized as follows:
\begin{enumerate}
\item 
The developed parametric schemes have demonstrated the ability to encode and reconstruct airfoils effectively, as analyzed comprehensively. 
The approach involves encoding existing airfoils into a lower-dimensional latent space, which serves as a geometric representation. 
The reconstructed airfoil geometries exhibit acceptable accuracy for airfoil parameterization purposes and are sufficiently smooth without requiring additional smoothing filters.

\item 
The completeness of the generated samples in the geometry space is evidenced by the distribution of the latent space of different datasets, where a well-trained model can capture the full characteristics of an airfoil and serve as variables for airfoil parameterization.
The combination of these characteristics enables the representation of both interpolation and extrapolation of known samples. The variance of the latent vector is indicative of the diverse geometry features of the synthesized airfoil based on the reference. This provides valuable insight into the variability of the generated airfoils, allowing for a deeper understanding of the synthesized airfoil geometries.

\item 
The proposed schemes enable the incorporation of geometric constraints during airfoil parameterization. 
The soft scheme generates airfoils with slight deviations from the expected geometric constraints, but exhibits higher efficiency and effectiveness in generating airfoils. 
The hard scheme generates airfoils with a broader range of geometric diversity and strictly satisfies the geometric constraints. Nevertheless, there is a low likelihood of generating abnormal airfoils, which can be readily mitigated by controlling the perturbation range of the variables.

\item 
The proposed schemes exhibit the ability to surpass the constraints of training data in the objective space, resulting in the generation of superior quality samples for random sampling and optimization design enhancement. 
The soft scheme achieves high efficiency and concentration near the reference point, making it suitable for local optimization. 
The hard scheme generates a broader range of objective space distribution, thereby providing more viable options for global optimization. 

\end{enumerate}

Our parametric approaches embody the geometric constraints of design requirements in an intuitive manner while leveraging a latent vector space with greater flexibility to achieve a more comprehensive description of the design space. Our experimental results attest to its potential as a fundamental design tool for airfoils, an efficient method for dataset sampling, and an optimization performance enhancer, thereby manifesting practical significance for engineering applications. In our future work, we aim to extend the scope of our research to encompass more intricate geometries and higher-dimensional problems with a view towards broadening the potential of our approach for real-world engineering applications.

% \appendix

% \section{Sample Appendix Section}
% \label{sec:sample:appendix}
% Lorem ipsum dolor sit amet, consectetur adipiscing elit, sed do eiusmod tempor section \ref{sec:sample1} incididunt ut labore et dolore magna aliqua. Ut enim ad minim veniam, quis nostrud exercitation ullamco laboris nisi ut aliquip ex ea commodo consequat. Duis aute irure dolor in reprehenderit in voluptate velit esse cillum dolore eu fugiat nulla pariatur. Excepteur sint occaecat cupidatat non proident, sunt in culpa qui officia deserunt mollit anim id est laborum.

\bibliographystyle{elsarticle-num-names} 
\bibliography{refs}





% \end{thebibliography}
\end{document}